\documentclass[twocolumn,showpacs,preprintnumbers,amsmath,amssymb]{revtex4}
\usepackage{graphicx}
\usepackage{amsmath}
\begin{document}

\title{Two-dimensional rocking ratchet for cold atoms}

\author{V. Lebedev and F. Renzoni}

\affiliation{Departement of Physics and Astronomy, University College London,
Gower Street, London WC1E 6BT, United Kingdom}

\date{\today}

\begin{abstract}
We investigate experimentally a two-dimensional rocking ratchet for cold atoms,
realized by using a driven three-beam dissipative optical lattice. AC forces are
applied in perpendicular directions by phase-modulating two of the lattice 
beams. As predicted by the general theory [S.~Denisov {\it et al.}, Phys. Rev. 
Lett. {\bf 100}, 224102 (2008)], we observe a rectification phenomenon unique
to high-dimensional rocking ratchets, as determined by two single-harmonic 
drivings applied in orthogonal directions. Also, by applying two bi-harmonic 
forces in perpendicular directions, we demonstrate the possibility of
generating a current in an arbitrary direction within the optical lattice plane.
\end{abstract}

\maketitle

\section{Introduction}

The ratchet effect 
\cite{comptes}
consists in the rectification of Brownian motion within a 
macroscopically flat potential, and corresponding generation of a current.
Due to the second principle of thermodynamics, directed motion in such a 
situation can only be obtained in out-of-equilibrium systems. Furthemore,
it is necessary to break all the system's symmetries which would otherwise
prevent directed motion \cite{flach00,flach01,super,denisov08}.
%%%
The ratchet effect is very general, as well exemplified by the variety of 
systems in which it has been demonstrated: from colloidal particles 
\cite{rousselet} and solid state devices \cite{linke,villegas,silva} 
to cold atoms in optical lattices 
\cite{schiavoni,phil,gommers05b,quasip,kastberg}, synthetic molecules
\cite{serreli} and granular gases \cite{granular}, just to name a few.
Among the different possible implementations of the ratchet effect, 
one-dimensional rocking ratchets have been studied in great detail.
In these ratchets, Brownian particles in an asymmetric periodic potential 
are driven out of equilibrium by a time-symmetric oscillating force.
The particles are set into directed motion due to the asymmetry of the 
potential landscape. The same effect can be obtained for a spatially
symmetric potential and a temporally asymmetric drive 
\cite{fabio,mahato,flach00,flach01}.  A typical choice for a time asymmetric 
drive is a bi-harmonic force, with the relative phase between harmonics 
determining the time-symmetry of the drive 
\cite{flach00,flach01,super,denisov08}. Mixing of harmonics with different
parity produces then directed motion through the spatially symmetric potential
\cite{fabio}.

Recently, the possibility of controlling the motion, via the ratchet effect,
within a 2D or a 3D structure has been attracting much attention. In this 
context, a 3D Brownian motor was demonstrated by using cold atoms in 
{\it undriven} optical lattices \cite{kastberg}. Stimulated by recent 
theoretical work \cite{sergey05,denisov08,olson}, in the present work
we consider  2D {\it ac driven} 
rocking ratchets for cold atoms as a model system to investigate rectification
phenomena unique to these higher dimensional systems, as produced by the 
interplay between drivings applied in orthogonal directions.
We demonstrate that directed motion can be obtained by two 
single-harmonic drivings applied in orthogonal directions, a rectification 
phenomenon unique to high-dimensional rocking ratchets. Furthemore, by applying two 
bi-harmonic drivings in orthogonal directions, we demonstrate the possibility 
of generating a current in an arbitrary direction within the optical lattice 
plane.  

\section{Experimental set-up}

\begin{figure}[hbtp]
\begin{center}
\includegraphics[height=1.5in]{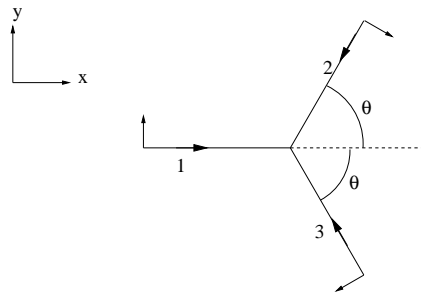}
\end{center}
\caption{
Lattice beam configuration for the 2D near-resonant optical lattice
used in this work. The intensity per lattice beam is $I_L=70$ mW/cm$^2$.
The detuning of the lattice fields from resonance with the 
D$_2$-line $F_g=2\to F_g=3$ atomic transition is
$\Delta =-15 \Gamma$. The angle $\theta$ is $\theta=60^{\circ}$.}
\label{fig:fig1}
\end{figure}

Our ratchet set-up consists of a dissipative 2D optical lattice generated by
three laser beams of equal intensity \cite{robi}. The lattice beam
configuration is shown in Fig.~\ref{fig:fig1}: three linearly polarized
travelling waves propagate in the $xy$ plane; their propagation directions
are separated by 120$^\circ$, and their linear polarizations are all in the
$xy$ plane. This beam configuration gives rise to a periodic 2D optical
lattice, with the potential minima arranged on a hexagonal lattice.
The use of just three beams to generate a 2D lattice has an important advantage
\cite{robi}: changes in the relative phases between the different laser beams 
do not produce a change in the topography of the optical potential, but only
result in a shift of the lattice as a whole. This allows us to introduce
rocking forces in the $x$- and $y$- directions by phase-modulating two 
of the lattice beams. The basic idea is to operate a moving optical lattice,
with time-dependent acceleration ${\bf a}$. Then, in the non-inertial reference
frame of the lattice, an inertial force ${\bf F} = - m {\bf a}$ is introduced, 
where $m$ is the atomic mass.  In the specific case considered here, 
if we modulate beams 2 and 3, and indicate the resulting
time-dependent phases by $\phi_2$ and $\phi_3$ respectively, we find that in 
the reference frame of the lattice an inertial force ${\bf F}$ appears with 
components equal to:
\begin{equation}
{\bf F} = (F_x,F_y) =-m (a_x,a_y) =  - \frac{m}{k} (\frac{\ddot{s}}{3}, \frac{\ddot{\delta}}{\sqrt{3}}) 
\label{eq:force}
\end{equation}
Here $a_x$, $a_y$ are the components of the lattice acceleration, 
$k$ the laser beam's wavevector, $s$ and $\delta$ are the sum and the 
difference of the phases of beams $1$ and $2$: $ s = \phi_2 + \phi_3$, 
$\delta = \phi_3 - \phi_2$.
In the experiment, the three lattice beams are obtained from the same laser 
beam. This laser beam is split into three beams, which are passed through three
acousto-optical modulators (AOM1, AOM2 and AOM3 for the beams 1,2 and 3,
respectively) driven by three RF generators at 80 MHz. The 
first-order diffracted beams are then used to generate the optical lattice.
Three additional phase-locked frequency generators are used to produce the 
rocking forces $F_x$ and $F_y$. One generator produces the signals
$V_1 = A_x \sin(\omega t)$ and $V_2 = A_y \sin(\omega t)$, while the other two generators produce 
$V_3=B_x\sin(2\omega t+\phi_x)$ and $V_4=B_y\sin( 2\omega t + \phi_y)$ with 
$\omega = 2\pi\cdot 50$ kHz for all experiments reported in this work. We 
then add the signals two by two, so to obtain:
\begin{subequations}  \label{eq:voltages}
\begin{eqnarray}
V_x &=&  A_x\sin(\omega t) + B_x \sin( 2\omega t + \phi_x)\\
V_y &=&  A_y\sin(\omega t) + B_y \sin( 2\omega t + \phi_y)~.
\end{eqnarray}
\end{subequations}
The sum $V_x+V_y$ is then used as frequency-modulation input signal for the 
RF generator driving AOM2, so to phase modulate lattice beam $2$. In the same
way, the difference $V_x-V_y$ is used to phase-modulate, via the RF
generator driving AOM3, the lattice beam $3$. According to Eq.~\ref{eq:force}
this produces the following ac forces:
\begin{subequations} \label{eq:forces2}
\begin{eqnarray}
F_x &=& -\frac{2m\omega f_0}{3k} 
        \left[ A_x\cos(\omega t) + 2 B_x\cos(2\omega t+\phi_x)\right]\\
F_y &=& -\frac{2m\omega f_0}{\sqrt{3} k} 
        \left[ A_y\cos(\omega t) + 2 B_y\cos(2\omega t+\phi_y)\right]
\end{eqnarray}
\end{subequations} 
where $f_0 =500$ kHz, as determined by the frequency modulation depth of 
the RF generators.

\section{Symmetry analysis}

Before presenting the findings of our experiments, we recall the essential elements of the
symmetry analysis for a two-dimensional rocking ratchet \cite{denisov08}.
Such an analysis will allow us to interpret the experimental results. Consider
a particle moving in a bi-dimensional potential $V(x,y)$ and driven by a
periodic zero-average ac force ${\bf F}(t)$ of period $T$. The relevant
symmetries for the directed transport through the potential are those
which reverse the sign of the momentum:
\begin{eqnarray}
S_1:&&~{\bf r} \to {\bf -r+r'},~~ t\to t+\tau \\
S_2:&&~{\bf r} \to {\bf r+\Lambda},~~ t\to -t+t'~,
\end{eqnarray}
where ${\bf r}'$, $\tau$, $\bf {\Lambda}$ and $t'$ are constants which
represent shifts in time and space.
If the system is invariant under $S_1$ and/or $S_2$, directed
transport is forbidden.
Whether  $S_1$, $S_2$ are symmetries of the system depends on the symmetry
properties of the potential and the driving. If the potential is symmetric,
i.e. $V(-{\bf r+r'})=V({\bf r})$, and the driving is {\it shift-symmetric},
i.e. ${\bf F}(t+T/2)=-{\bf F}(t)$, then $S_1$ is a symmetry of the system.
Moreover, in the Hamiltonian limit $S_2$ holds if the driving is symmetric
under time-reversal, i.e. ${\bf F}(-t+t')={\bf F}(t)$.

In the following, the effect of the application of different ac forces 
will be studied by analyzing the breaking of the relevant symmetries.

\section{Experimental results}

The experimental procedure is the following. We cool and trap $^{87}$Rb atoms 
in a magneto-optical trap (MOT). Once the trap is loaded, we turn off the MOT
beams and magnetic field, and load the atoms in the aforedescribed 2D optical
lattice. The ac drivings are then turned on by increasing linearly with 
time their amplitudes from zero to the wanted value in $1$ ms. The motion of
the atoms in the driven lattice is then studied with the help of a CCD camera 
using absorption imaging. This allows us to determine the position of the 
atomic cloud center of mass at a given instant.

\subsection{1D ratchet effect}

\begin{figure}[hbtp]
\begin{center}
\includegraphics[width=3.in]{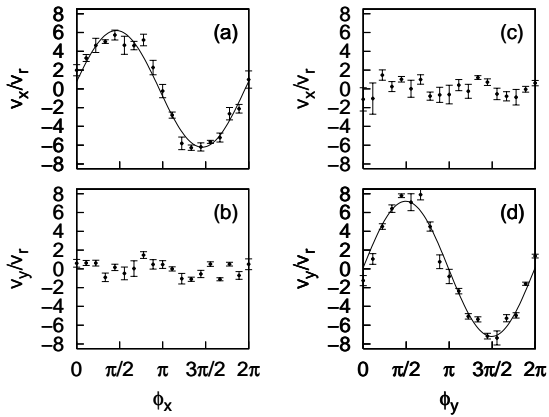}
\end{center}
\caption{
Experimental results for the $x$- and $y$- components of the center-of-mass
velocity of the atomic cloud as a function of the phase between the driving
harmonics. The left column (plots (a) and (b)) refers to a bi-harmonic driving
along $x$.  The right column (plots (c) and (d)) refers to a bi-harmonic
driving along $y$. The parameters of the driving (see Eq.~\protect\ref{eq:forces2})
are: $A_x=1.5$V, $B_x=2.0$V, $A_y=B_y=0$V for (a) and (b), and 
$A_x=B_x=0$V, $A_y=1.5$V, $B_y=2.0$V for (c) and (d). The solid lines are the 
best fit of the data with the function
$v_{x,y} = v_{\rm max}^{x,y}\sin(\phi_{x,y}+\phi_0^{x,y})$.
}
\label{fig:fig_exp1}
\end{figure}

In the first set of measurements, we apply a bi-harmonic driving along the 
$x$- {\it or} the $y$-direction, and monitor the atomic center-of-mass (CM) 
motion.
Results for the $x-$ and $y-$ components of the CM velocity are reported in 
Fig.~\ref{fig:fig_exp1} as a function of the phase difference between the 
driving harmonics. The data of Fig.~\ref{fig:fig_exp1}(a,b) refer to a 
bi-harmonic driving along the $x$-direction, while the data of 
Fig.~\ref{fig:fig_exp1}(c,d) refer to a driving along the $y$ direction. These
measurements show that a bi-harmonic driving in the $x$- (respectively, $y-$)
direction leads to a ratchet effect only in that direction, with the current 
generated showing a sinusoidal dependence on the phase between driving 
harmonics.

We notice that for a bi-harmonic driving applied in one direction only, 
the symmetry analysis reduces to the one for a 1D rocking ratchet
\cite{flach00,phil,gommers05b}. This is consistent with
our results of Fig.~\ref{fig:fig_exp1}, with a current generated in the 
direction of the driving only.

\subsection{Split bi-harmonic driving}

\begin{figure}[hbtp]
\begin{center}
\includegraphics[width=3.in]{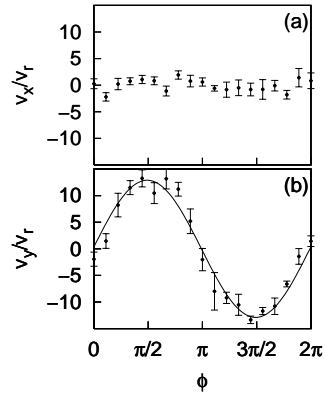}
\end{center}
\caption{Experimental results for the $x$- and $y$- components of the 
center-of-mass velocity of the atomic cloud as a function of the phase 
between the driving harmonics. The driving consists of a single harmonic 
driving, with frequency $\omega$, along $x$ and an additional single harmonic
driving, with frequency $2\omega$ and relative phase $\phi$, along $y$.
The parameters of the driving (see Eq.\protect\ref{eq:forces2})
are: $A_x=1.5$V, $B_x=0$V, $A_y=0$V, $B_y=2.0$V.  The solid line is the
best fit of the data with the function 
$v_y = v_{\rm max}\sin(\phi+\phi_0)$.
}
\label{fig:fig_exp2}
\end{figure}

Consider now the case of simultaneous driving along the $x$- and $y$- 
directions. We examine first the case of a single harmonic 
driving, of frequency $\omega$, along the $x$-direction and an additional 
single harmonic driving, with frequency $2\omega$ and relative phase $\phi$,
along $y$ ("split bi-harmonic driving"). 
The results of our measurements for this case are reported in 
Fig.~\ref{fig:fig_exp2}. For this configuration a phase-dependent current 
is generated along the $y$ direction, while no current is generated, within 
the experimental error, in the $x$ direction. Such a generation of a current
can be understood in the framework of the symmetry analysis \cite{denisov08}.
We consider the Hamiltonian limit, with a phase-shift accounting for weak 
dissipation. This is appropriate for a weakly-dissipative rocking ratchet, as 
the ones realized by using near-resonant optical lattices \cite{gommers05b,brown}. 
The driving is of the form 
\begin{equation}
{\bf F} = \hat{\bf x}F_x^\circ \cos(\omega t) + 
          \hat{\bf y}F_y^\circ \cos(2\omega t+\phi),
\end{equation}
with the period of the driving equal to $T = 2\pi/\omega$. As we have 
$V(x,y)=V(-x,y)$ and $F_x(t+T/2) = -F_x(t)$, directed transport along
$x$ is forbidden by symmetry. On the other hand, the $y$-component of the
driving force is not shift-symmetric as $F_y(t+T/2)\neq -F_y(t)$. Thus,
transport along $y$ is not forbidden by symmetry. Such a transport is then 
controlled by $S_2$, which is realized for $\phi=n\pi$, with $n$ integer.
The above symmetry analysis explains the observed generation of a 
phase-dependent sine-like current along $y$, while no current is generated 
along $x$. We obtained analogous results for the reversed situation: by 
applying a single harmonic driving along $y$ with frequency $\omega$, and a 
single harmonic driving along $x$ with frequency $2\omega$ and relative 
phase $\phi$, we observed the generation of a phase-dependent sine-like 
current along the $x$-direction.

These results constitute a first experimental demonstration of a rectification
phenomenon unique to high dimensional rocking ratchets, as predicted
by the symmetry analysis of Denisov {\it et al} \cite{denisov08}. 
Furthermore, these results show that for the
considered lattice geometry the $x$- and $y$-directions are coupled under
the influence of ac drivings. That is, the application of two single-harmonic 
drivings along the $x$- and $y$- directions, which taken one-by-one would not 
break the relevant symmetries and would therefore not induce any directed 
motion, can break the symmetries of the system and lead to directed motion.
Under the action of these forces the atoms explore orbits in 2D \cite{denisov08}
while generating an average drift in the direction of the driving at the frequency $2\omega$. 

\subsection{Control of directed motion in 2D}

We now examine the implementation of a 2D rocking ratchet in which 
directed motion can be controlled in the $xy$ plane.
We consider the case of two bi-harmonic drivings applied simultaneously, one
in the $x$- and one in the $y$-direction. The results presented so far - see 
Fig.~\ref{fig:fig_exp1} and related discussion - showed that by applying a single
bi-harmonic driving along the $x$ or $y$ direction we can generate a current in
the direction of the driving. The issue addressed now is whether by applying 
simultaneously two bi-harmonic drivings, in the $x$ and $y$ directions 
respectively, it is possible to generate a current in an arbitrary direction in 
the $xy$ plane.

\begin{figure}[hbtp]
\begin{center}
\includegraphics[width=3.in]{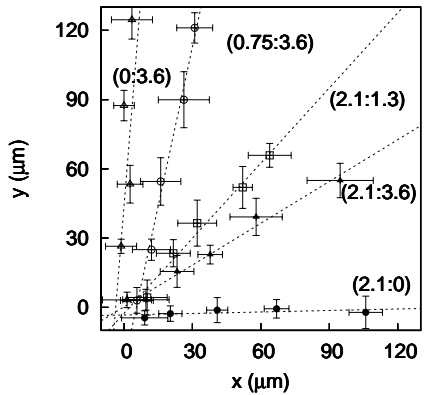}
\end{center}
\caption{
Position of the atomic cloud center-of-mass at different instants. The 
different data sets correspond to different relative amplitudes of
$F_x$ and $F_y$, and are labelled by $(V_x^{pp}/\sqrt{3}:V_y^{pp})$,
where $V^{pp}_x$ (respectively, $V^{pp}_y$) is the peak-to-peak amplitude
of the signal - see Eq.~\ref{eq:voltages} - leading to the generation of 
the ac force in the $x$ (respectively, $y$) direction.  As from 
Eq.~\ref{eq:forces2}, the ratio $(V_x^{pp}/\sqrt{3})/V_y^{pp})$ is 
proportional to the forces amplitude ratio $F_x/F_y$. For each data set, 
the point closer to the origin corresponds to the first image, taken 
immediately after the ac drivings were ramped up. The other points correspond 
to images taken at intervals of $0.5$ ms. The lines are the best fit of
the data with a linear function.
}
\label{fig:fig_exp3}
\end{figure}

We consider two-biharmonic driving forces $F_x$, $F_y$ of the form specified
by Eq.~\ref{eq:forces2}. We take the two forces to have the same temporal
dependence by choosing the same relative phases betwen harmonics:
$\phi_x=\phi_y=\pi/2$ and the same ratio between harmonic amplitudes:
$A_x/B_x=A_y/B_y=3/4$. We notice that the choice of the values for the
phases is such to break all the relevant symmetries which would otherwise
inhibit the generation of a current. This is confirmed by our results of
Fig.~\ref{fig:fig_exp1}: for this choice of phases and ratio between harmonic
amplitudes, $F_x$ and $F_y$ individually can generate
directed motion along $x$ and $y$ respectively.

In the experiment, we apply $F_x$ and $F_y$ simultaneously and monitor
the resulting motion of the atomic center-of-mass. The results of our
measurements are reported in Fig.~\ref{fig:fig_exp3}, where the
center-of-mass of the atomic-cloud is plotted at different instants.
The different series of measurements correspond to different relative
weights of $F_x$ and $F_y$, as obtained by varying the overall
amplitudes of $F_x$ and $F_y$ while keeping their temporal dependence
unchanged. Our measurements show that the atoms are set into directed motion
as a result of the combined action of the drivings in the $x$- and $y$-
directions. And it is possible, by choosing appropriately the relative weights
of the drivings in the $x$ and $y$ directions, to generate a current in
a wanted direction. 

It is interesting to stress some characteristics of the generation of a
current in a 2D rocking ratchet, as from Fig.~\ref{fig:fig_exp3}. The 
generated current should not be interpreted as consisting of a 
$x$-component $v_x$ generated by $F_x$ only and by a $y$-component 
$v_y$ generated by $F_y$ only. This because there is a cross-coupling 
between orthogonal directions, as best exemplified by our results of a 
generation of a current following two single-harmonic drivings in the 
$x$- and $y$-directions respectively - see Fig.~\ref{fig:fig_exp2}.
This clearly produces a complicated dependence of the directions of 
the obtained current on the amplitudes of $F_x$ and $F_y$. Moreover, 
we notice that even for a 1D ratchet the current amplitude is a 
non-monotonic function of the driving amplitude \cite{phil}. Therefore,
even neglecting cross-couplings, we do not expect the direction of
the current to vary monotonically with the ratio of the amplitudes 
between the bi-harmonic drivings in the $x$- and $y$-directions.

\section{Conclusions}

In conclusion, we investigated a two-dimensional rocking ratchet for cold atoms
realized using ac driven dissipative optical lattices. 
In these optical lattices the excess energy introduced by
the driving is dissipated by the friction force associated to the Sisyphus
cooling mechanism \cite{robi}, and it is removed from the system by the
scattered photons.  We observed a rectification phenomenon unique 
to high-dimensional rocking ratchets, as determined by two single-harmonic drivings applied 
in orthogonal directions, and demonstrated the possibility of generating 
a current in an arbitrary direction within the optical lattice
plane.

The set-up demonstrated in this work will also allow one to investigate
several other phenomena specific to high-dimensional rocking ratchets, as
the generation and control of vorticity \cite{denisov08} or the transverse
rectification of fluctuations resulting from orthogonally applied
dc and ac drivings \cite{olson}. The results presented in this work are also of 
relevance to fluxtronic devices, in which the motion of flux quanta
within a 2D spatially symmetric landscape is controlled by time-asymmetric ac 
fields \cite{ooi}.

We thank the Royal Society and the Leverhulme Trust for financial support.


\begin{thebibliography}{99}
\bibitem{comptes}
A.~Ajdari and J.~Prost, C.R. Acad. Sci. Paris {\bf 315}, 1635 (1992);
%\bibitem{magnasco}
M.O.~Magnasco, Phys. Rev. Lett. {\bf 71}, 1477 (1993);
%\bibitem{adjari}
A.~Adjari {\it et al.}, J. Phys. I (France)
{\bf 4}, 1551 (1994);
%\bibitem{bartussek}
R.~Bartussek, P.~H\"anggi and J.G.~Kissner, Europhys. Lett. {\bf 28},
459 (1994);
%\bibitem{doering}
C.R.~Doering, W.~Horsthemke, and J. Riordan, Phys. Rev. Lett. {\bf 72},
2984 (1994);
%\bibitem{astumian}
R.D.~Astumian, Science {\bf 276}, 917 (1997);
%\bibitem{reimann}
P.~Reimann, Phys. Rep. {\bf 361}, 57 (2002);
%\bibitem{rmp09}
P. H\"anggi and F.~Marchesoni, Rev. Mod. Phys. {\bf 81}, 387 (2009).
\bibitem{flach00}
S.~Flach, O.~Yevtushenko and Y.~Zolotaryuk, Phys. Rev. Lett. {\bf 84},
2358 (2000);
\bibitem{flach01}
O.~Yevtushenko {\it et al.},
Europhys. Lett. {\bf 54}, 141 (2001).
\bibitem{super}
P.~Reimann, Phys. Rev. Lett. {\bf 86}, 4992 (2001).
\bibitem{denisov08}
S.~Denisov {\it et al.}, Phys. Rev. Lett. {\bf 100}, 224102 (2008).
\bibitem{rousselet}
J.~Rousselet {\it et al.}, Nature {\bf 370}, 446 (1994).
\bibitem{linke}
H.~Linke {\it et al.}, Science {\bf 286}, 2314 (1999).
\bibitem{villegas}
J.E.~Villegas {\it et al.}, Science {\bf 302}, 1188 (2003).
\bibitem{silva}
C.C.~de Souza Silva {\it et al.}, Nature {\bf 440}, 651 (2006).
\bibitem{schiavoni}
M. Schiavoni {\it et al.},
Phys. Rev. Lett. {\bf 90}, 094101 (2003).
\bibitem{phil}
P.H.~Jones, M.~Goonasekera, and F.~Renzoni,
Phys. Rev. Lett. {\bf 93}, 073904 (2004).
% \bibitem{gommers05a}
% R.~Gommers {\it et al.}, Phys. Rev. Lett. {\bf 94}, 143001 (2005).
\bibitem{gommers05b}
R.~Gommers, S.~Bergamini, and F.~Renzoni,
Phys. Rev. Lett. {\bf 95}, 073003 (2005).
\bibitem{quasip}
R.~Gommers, S.~Denisov and F.~Renzoni, Phys. Rev. Lett. {\bf 96}, 240604 (2006);
R.~Gommers, M.~Brown, and F.~Renzoni, Phys. Rev. A {\bf 75}, 053406 (2007).
\bibitem{kastberg}
P.~Sjolund {\it et al.}, Phys. Rev. Lett. {\bf 96}, 190602 (2006);
P.~Sjolund {\it et al.}, Eur. Phys. J. D {\bf 44}, 381 (2007).
\bibitem{serreli}
V.~Serreli {\it et al.}, Nature {\bf 445}, 523 (2007).
\bibitem{granular}
D.~van der Meer {\it et al.}, Phys. Rev. Lett.
{\bf 92}, 184301 (2005).
\bibitem{fabio}
F.~Marchesoni, Phys. Lett. A {\bf 119}, 221 (1986).
\bibitem{mahato}
M.C.~Mahato and A.M.~Jayannavar, Phys. Lett. A {\bf 209}, 21 (1995);
% \bibitem{chialvo}
D.R.~Chialvo and M.M. Millonas, Phys. Lett. A {\bf 209}, 26 (1996);
% \bibitem{dykman}
M.I.~Dykman {\it et al.},  Phys. Rev.  Lett. {\bf 79}, 1178 (1997);
% \bibitem{goychuk}
I.~Goychuk and P.~H\"anggi,  Europhys. Lett. {\bf 43}, 503 (1998).
\bibitem{sergey05}
S.~Savel'ev, Phys. Rev. B {\bf 71}, 214303 (2005).
\bibitem{olson}
C.~Reichhardt, C.J.~Olson, and M.B.~Hastings, 
Phys. Rev. Lett. {\bf 89}, 024101 (2002).
% \bibitem{nori04}
% S.~Savel'ev {\it et al.}, Europhys. Lett. {\bf 67}, 179 (2004);
% M.~Borromeo, and F.~Marchesoni, Chaos {\bf 15}, 026110 (2005);
% M.~Borromeo, S.~Giusepponi, and F.~Marchesoni, Phys. Rev. E {\bf 74}, 031121 
% (2006).
\bibitem{robi}
G.~Grynberg and C.~Mennerat-Robilliard, Phys. Rep. {\bf 355}, 335 (2001).
\bibitem{brown}
M.~Brown and F.~Renzoni, Phys. Rev. A {\bf 77}, 033405 (2008).
\bibitem{ooi}
S.~Savel'ev and F.~Nori, Nat. Mater. {\bf 1}, 179 (2002);
D.~Cole {\it et al.}, Nat. Mater. {\bf 5}, 305 (2006);
S.~Ooi {\it et al}, Phys. Rev.  Lett. {99}, 207003 (2007).
\end{thebibliography}
\end{document}